\documentclass{elsart}

\usepackage{epsfig}
\usepackage{amssymb}

\begin{document}

\begin{frontmatter}
\title{Probabilistic implementation of Hadamard and Unitary gates}
\author{Wei Song\corauthref{cor}},
\corauth[cor]{Corresponding author.} \ead{wsong@mars.ahu.edu.cn}
\author{Ming Yang},
\author{Zhuo-Liang Cao}
\ead{zlcao@mars.ahu.edu.cn}
\address{School of Physics {\&} Material Science, Anhui University, Hefei, 230039, P.
R. of China}
\begin{abstract}
We show that the Hadamard and Unitary gates could be implemented
by a unitary evolution together with a measurement for any unknown
state chosen from a set $A = \left\{ {\left| {\Psi _i }
\right\rangle ,\left| {\bar {\Psi }_i } \right\rangle }
\right\}\left( {i = 1,2} \right)$ if and only if $\left| {\Psi _1
} \right\rangle ,\left| {\Psi _2 } \right\rangle ,\left| {\bar
{\Psi }_1 } \right\rangle ,\left| {\bar {\Psi }_2 } \right\rangle
$ are linearly independent. We also derive the best transformation
efficiencies.
\end{abstract}


\end{frontmatter}

\ Manipulation and extraction of quantum information are important
tasks in building quantum computer. Unlike classical information
there are several limitations on the basic operations that one can
perform on quantum information. Linearity of quantum mechanics
unveils that we cannot duplicate an unknown quantum state
accurately\cite{Wootters:1982}. This has been proven by Wootter
and Zurek\cite{Wootters:1982} and Dieks\cite{Dieks:1982} which
called the quantum no-cloning theorem. Though exact cloning is not
possible, in the literature various cloning machines have been
proposed
\cite{Buzek:1996,Gisin:1997,Bruss:1998,Werner:1998,Keyl:1999,Han:2002,Fan:2002,Duan:1998,Duan:1999,Buzek:1999,Chefles:1999,Bruss:2000,Cerf:2000,Fiurasek:2002,Braunstein:2001,Buzek:1997,Zhang:2000,Horodecki:1,Fiurasek:1,Cao:2004}
which operate either in a deterministic or probabilistic way.
Corresponding to the quantum no-cloning theorem, Pati and
Braunstein \cite{Pati:2000} demonstrated that the linearity of
quantum mechanics also forbids one to delete one unknown state
ideally against a copy \cite{Pati:2000}, which is called the
quantum no-deleting principle. Notably, Zurek \cite{Zurek:2000}
further verified the existence of limitations on cloning and
deleting completely an unknown state and pointed out the
importance of studying approximate and probabilistic deletion
corresponding to cloners \cite{Buzek:1996,Gisin:1997,Duan:1998}.
And some probabilistic and state-dependent deleting machines have
been established \cite{Feng:2001,Qiu:2002,Song:2004}. In addition,
it is found that one cannot flip a spin of unknown polarization,
because the flip operator $V$ defined as

\begin{equation}
\label{eq1}
V\left| \Psi \right\rangle = \left| \bar {\Psi } \right\rangle
\end{equation}

\noindent
is not unitary but anti-unitary, where if qubit $\left| \Psi \right\rangle =
\alpha \left| 0 \right\rangle + \beta \left| 1 \right\rangle $ belongs to
two-dimensional Hilbert space, then its opposite direction is defined as
$\left| \bar {\Psi } \right\rangle = \beta ^\ast \left| 0 \right\rangle -
\alpha ^\ast \left| 1 \right\rangle $, but there is no physical operation
capable of implementing such a transformation. This is called the quantum
no-complementing principle
\cite{Buzek:1999,Gisin:1999,Pati:2002}. It is
also shown that if we are given an arbitrary state $\left| \Psi
\right\rangle \in H^2$ of an unknown qubit and a blank state $\left| \Sigma
\right\rangle \in H^2$, there does not exist an isometric operator $U$ such
that

\begin{equation}
\label{eq2}
U\left| \Psi \right\rangle \left| \Sigma \right\rangle = \left| \Psi
\right\rangle \left| \bar {\Psi } \right\rangle ,
\end{equation}

\noindent
holds \cite{Pati:2002}, which is called the quantum no-anti-cloning
property. Lately, Pati considered the question if we are given an unknown
qubit pointed in some arbitrary direction $n$ in a state $\left| \Psi
\right\rangle $ or in the direction $ - n$ in a state$\left| \bar {\Psi }
\right\rangle $ can we design a logic gate that will transform these inputs
as follows \cite{Pati:2002}:

\[
\left| \Psi \right\rangle \to \frac{1}{\sqrt 2 }\left( {\left| \Psi
\right\rangle + \left| \bar {\Psi } \right\rangle } \right),
\]

\begin{equation}
\label{eq3}
\left| \bar {\Psi } \right\rangle \to \frac{1}{\sqrt 2 }\left( {\left| \Psi
\right\rangle - \left| \bar {\Psi } \right\rangle } \right),
\end{equation}

\noindent
where $\left| \Psi \right\rangle $ is an unknown state $\left| \Psi
\right\rangle = \alpha \left| 0 \right\rangle + \beta \left| 1 \right\rangle
\in H^2$, with $\alpha $ and $\beta $ being unknown complex numbers and
$\left| \alpha \right|^2 + \left| \beta \right|^2 = 1$, and $\left| \bar
{\Psi } \right\rangle $ is the complement of $\left| \Psi \right\rangle $.
They proved that there is no universal Hadamard gate defined by (\ref{eq3}) for an
unknown qubit that will create an equal superposition of the original state
$\left| \Psi \right\rangle $ and its complement state $\left| \bar {\Psi }
\right\rangle $. They also show that it is not possible to design a unitary
transformation that will create an unequal surperposition of the original
qubit with its complement which will transform the inputs as follows
\cite{Pati:2002}.

\[
\left| \Psi \right\rangle \to a\left| \Psi \right\rangle + b\left| \bar
{\Psi } \right\rangle ,
\]

\begin{equation}
\label{eq4}
\left| \bar {\Psi } \right\rangle \to b^\ast \left| \Psi \right\rangle -
a^\ast \left| \bar {\Psi } \right\rangle ,
\end{equation}

\noindent
where $a,b$ are known complex numbers and $\left| a \right|^2 + \left| b
\right|^2 = 1$.

Analogous to the probabilistic cloning we will prove that the above
transformation defined by Eq. (\ref{eq3}) and Eq. (\ref{eq4}) could be implemented by a
unitary evolution together with a measurement if $\left| {\Psi _i }
\right\rangle $ and $\left| {\bar {\Psi }_i } \right\rangle $ are chosen
from a linearly independent set. Firstly we will consider how to implement
the transformation defined by Eq. (\ref{eq3}). The result is the following theorem:

Theorem 1. There exists a unitary operator $U$ such that for any unknown
state chosen from a set $A = \left\{ {\left| {\Psi _i } \right\rangle
,\left| {\bar {\Psi }_i } \right\rangle } \right\}\left( {i = 1,2} \right)$

\begin{equation}
\label{eq5}
U\left( {\left| {\Psi _i } \right\rangle \left| {P_0 } \right\rangle }
\right) = \sqrt {\gamma _i } \frac{1}{\sqrt 2 }\left( {\left| {\Psi _i }
\right\rangle + \left| {\bar {\Psi }_i } \right\rangle } \right)\left| {P_0
} \right\rangle + \sum\limits_{j = 1}^2 {a_{ij} \left| {\Phi _A^{\left( j
\right)} } \right\rangle } \left| {P_j } \right\rangle ,
\end{equation}

\begin{equation}
\label{eq6}
U\left( {\left| {\bar {\Psi }_i } \right\rangle \left| {P_0 } \right\rangle
} \right) = \sqrt {\delta _i } \frac{1}{\sqrt 2 }\left( {\left| {\Psi _i }
\right\rangle - \left| {\bar {\Psi }_i } \right\rangle } \right)\left| {P_0
} \right\rangle + \sum\limits_{j = 1}^2 {b_{ij} \left| {\bar {\Phi
}_A^{\left( j \right)} } \right\rangle } \left| {P_j } \right\rangle ,
\end{equation}

\noindent
for some real numbers $\gamma _i \ge 0$, $\delta _i \ge 0$, $a_{ij} \ge 0$
and $b_{ij} \ge 0$ with $i,j = 1,2$, if and only if $\left| {\Psi _1 }
\right\rangle ,\left| {\Psi _2 } \right\rangle ,\left| {\bar {\Psi }_1 }
\right\rangle ,\left| {\bar {\Psi }_2 } \right\rangle $ are linearly
independent, where $\left| {P_0 } \right\rangle ,\left| {P_1 } \right\rangle
,\left| {P_2 } \right\rangle $ denoting the probe states are orthonormal,
and the states $\left| {\Phi _A^{\left( j \right)} } \right\rangle $'s of
system A are normalized but unnecessarily orthogonal, $\left| {\bar {\Psi
}_i } \right\rangle $ and $\left| {\bar {\Phi }_A^{\left( j \right)} }
\right\rangle $ are complement of $\left| {\Psi _i } \right\rangle $ and
$\left| {\Phi _A^{\left( j \right)} } \right\rangle $ respectively. To prove
the existence of the unitary operator $U$ described by Eq. (\ref{eq5}) and Eq. (\ref{eq6}),
we first notice the following lemma \cite{Duan:1998}.

Lemma 1. If two sets of states $\left| {\phi _1 } \right\rangle ,\left|
{\phi _2 } \right\rangle ,...,\left| {\phi _n } \right\rangle $, and $\left|
{\tilde {\phi }_1 } \right\rangle ,\left| {\tilde {\phi }_2 } \right\rangle
,...,\left| {\tilde {\phi }_n } \right\rangle $ satisfy the condition

\begin{equation}
\label{eq7}
\left\langle {\phi _i } \right|\left. {\phi _j } \right\rangle =
\left\langle {\tilde {\phi }_i } \right|\left. {\tilde {\phi }_j }
\right\rangle ,\left( {i = 1,2,...,n;j = 1,2,...,n} \right),
\end{equation}

\noindent
there exists a unitary operator $U$ to make $U\left| {\phi _i }
\right\rangle = \left| {\tilde {\phi }_i } \right\rangle ,(i = 1,2,...,n)$.

We should also need to know the following fact: The linear independence of
$\left\{ {\left| {\Psi _i } \right\rangle ,\left| {\bar {\Psi }_i }
\right\rangle } \right\}\left( {i = 1,2} \right)$ implies that $\left\{
{\left| {\Psi _i } \right\rangle } \right\}\left( {i = 1,2} \right)$ and
$\left\{ {\left| {\bar {\Psi }_i } \right\rangle } \right\}\left( {i = 1,2}
\right)$ are also of linear independence.

Now we prove the above theorem by using the lemma.

The $2\times 2$ inter-inner-products of Eq. (\ref{eq5}) and Eq. (\ref{eq6}) yield the matrix
equation

\begin{equation}
\label{eq8}
X^{(\ref{eq1})} = \sqrt \Gamma X^{\left( 2 \right)}\sqrt {\Gamma ^ + } + AA^ + ,
\end{equation}

\begin{equation}
\label{eq9}
X^{(\ref{eq3})} = \sqrt \Lambda X^{\left( 4 \right)}\sqrt {\Lambda ^ + } + BB^ + ,
\end{equation}

\noindent
where the $2\times 2$ matrixes $A = \left[ {a_{ij} } \right]$,$B = \left[
{b_{ij} } \right]$, $X^{\left( 1 \right)} = \left[ {\left\langle {\Psi _i }
\mathrel{\left| {\vphantom {{\Psi _i } {\Psi _j }}} \right.
\kern-\nulldelimiterspace} {\Psi _j } \right\rangle } \right]$, $X^{\left( 2
\right)} = \left[ {\frac{1}{2}\left( {\left\langle {\Psi _i }
\mathrel{\left| {\vphantom {{\Psi _i } {\Psi _j }}} \right.
\kern-\nulldelimiterspace} {\Psi _j } \right\rangle + \left\langle {\Psi _i
} \mathrel{\left| {\vphantom {{\Psi _i } {\bar {\Psi }_j }}} \right.
\kern-\nulldelimiterspace} {\bar {\Psi }_j } \right\rangle + \left\langle
{\bar {\Psi }_i } \mathrel{\left| {\vphantom {{\bar {\Psi }_i } {\Psi _j }}}
\right. \kern-\nulldelimiterspace} {\Psi _j } \right\rangle + \left\langle
{\bar {\Psi }_i } \mathrel{\left| {\vphantom {{\bar {\Psi }_i } {\bar {\Psi
}_j }}} \right. \kern-\nulldelimiterspace} {\bar {\Psi }_j } \right\rangle }
\right)} \right]$, $X^{\left( 3 \right)} = \left[ {\left\langle {\bar {\Psi
}_i } \mathrel{\left| {\vphantom {{\bar {\Psi }_i } {\bar {\Psi }_j }}}
\right. \kern-\nulldelimiterspace} {\bar {\Psi }_j } \right\rangle }
\right]$ and $X^{\left( 4 \right)} = \left[ {\frac{1}{2}\left( {\left\langle
{\Psi _i } \mathrel{\left| {\vphantom {{\Psi _i } {\Psi _j }}} \right.
\kern-\nulldelimiterspace} {\Psi _j } \right\rangle - \left\langle {\Psi _i
} \mathrel{\left| {\vphantom {{\Psi _i } {\bar {\Psi }_j }}} \right.
\kern-\nulldelimiterspace} {\bar {\Psi }_j } \right\rangle - \left\langle
{\bar {\Psi }_i } \mathrel{\left| {\vphantom {{\bar {\Psi }_i } {\Psi _j }}}
\right. \kern-\nulldelimiterspace} {\Psi _j } \right\rangle + \left\langle
{\bar {\Psi }_i } \mathrel{\left| {\vphantom {{\bar {\Psi }_i } {\bar {\Psi
}_j }}} \right. \kern-\nulldelimiterspace} {\bar {\Psi }_j } \right\rangle }
\right)} \right]$. The diagonal efficiency matrix $\Gamma $ is defined by
$\Gamma = diag\left( {\gamma _1 ,\gamma _2 } \right)$, hence $\sqrt \Gamma =
\sqrt {\Gamma ^ + } = diag\left( {\sqrt {\gamma _1 } ,\sqrt {\gamma _2 } }
\right)$. Lemma 1 shows that if Eq. (\ref{eq8}) and Eq. (\ref{eq9}) is satisfied with a
diagonal positive-definite matrix $\Gamma $ and $\Lambda $, the unitary
evolution (\ref{eq5}) and (\ref{eq6}) can be realized in physics. Let us prove the existing
of Eq. (\ref{eq8}) firstly. Following the Lemma 2 in Ref. \cite{Duan:1998},
since the states $\left| {\Psi _1 } \right\rangle ,\left| {\Psi _2 }
\right\rangle $ are linearly independent, then the matrix $X^{\left( 1
\right)}$ is positive definite. From the fact we know $\left\{ {\left| {\bar
{\Psi }_i } \right\rangle } \right\}\left( {i = 1,2} \right)$ is linearly
independent. As a consequence, one can show that the matrix $X^{\left( 2
\right)}$ is positive definite. From continuity, for small enough but
positive $\gamma _i $, the matrix $X^{\left( 1 \right)} - \sqrt \Gamma
X^{\left( 2 \right)}\sqrt {\Gamma ^ + } $ is also positive definite.
Therefore there is unitary matrix $V$ such that:

\begin{equation}
\label{eq10}
V^ + \left( {X^{\left( 1 \right)} - \sqrt \Gamma X^{\left( 2 \right)}\sqrt
{\Gamma ^ + } } \right)V = diag\left( {m_1 ,m_2 } \right),
\end{equation}

\noindent
for some real numbers $m_i > 0\left( {i = 1,2} \right)$. Now we choose

\begin{equation}
\label{eq11}
A = Vdiag\left( {\sqrt {m_1 } ,\sqrt {m_2 } } \right)V^ + .
\end{equation}

Equation (\ref{eq8}) is thus satisfied with a diagonal positive-definite efficiency
matrix $\Gamma $. Within the same approach of verify Eq. (\ref{eq8}), we could prove
that Eq. (\ref{eq9}) could also be satisfied with a diagonal positive-definite
efficiency matrix $\Lambda $. In general we have $\Gamma \ne \Lambda $. Thus
we complete the proof of Theorem 1. Actually we may find that if $\left|
{\Psi _1 } \right\rangle ,\left| {\Psi _2 } \right\rangle $ are linearly
dependent, then the matrix $X^{\left( 1 \right)}$ is only positive-
semidefinite. With a diagonal positive-definite matrix $\Gamma $, in
general, $X^{\left( 1 \right)} - \sqrt \Gamma X^{\left( 2 \right)}\sqrt
{\Gamma ^ + } $ is no longer a positive-semidefinite matrix. But the matrix
$AA^ + $ is positive-semidefinite. So Eq. (\ref{eq8}) could not be satisfied. With
similar procedure, we could prove that if $\left| {\bar {\Psi }_1 }
\right\rangle ,\left| {\bar {\Psi }_2 } \right\rangle $ are linearly
dependent, then Eq. (\ref{eq9}) also could not be satisfied. These show that the
Hadamard gate defined in Eq. (\ref{eq3}) could not be probabilistically realized by
any unitary reduction operation for any unknown state chosen from the
linearly dependent states $\left| {\Psi _1 } \right\rangle ,\left| {\Psi _2
} \right\rangle ,\left| {\bar {\Psi }_1 } \right\rangle ,\left| {\bar {\Psi
}_2 } \right\rangle $.

In the following, we derive the best possible efficiencies able to be
attained. The general unitary evolution of the system AP can be decomposed
as

\begin{equation}
\label{eq12}
U\left( {\left| {\Psi _i } \right\rangle \left| {P_0 } \right\rangle }
\right) = \sqrt {\gamma _i } \frac{1}{\sqrt 2 }\left( {\left| {\Psi _i }
\right\rangle + \left| {\bar {\Psi }_i } \right\rangle } \right)\left|
{P^{\left( i \right)}} \right\rangle + \sqrt {1 - \gamma _i } \left| {\Phi
_{AP}^{\left( i \right)} } \right\rangle ,
\end{equation}

\begin{equation}
\label{eq13}
U\left( {\left| {\bar {\Psi }_i } \right\rangle \left| {P_0 } \right\rangle
} \right) = \sqrt {\delta _i } \frac{1}{\sqrt 2 }\left( {\left| {\Psi _i }
\right\rangle - \left| {\bar {\Psi }_i } \right\rangle } \right)\left|
{P^{\left( i \right)}} \right\rangle + \sqrt {1 - \delta _i } \left| {\bar
{\Phi }_{AP}^{\left( i \right)} } \right\rangle ,\left( {i = 1,2} \right),
\end{equation}

\noindent
where $\left| {P_0 } \right\rangle $, and $\left| {P^{\left( i \right)}}
\right\rangle $ are normalized states of the probe P and $\left| {\Phi
_{AP}^{\left( 1 \right)} } \right\rangle $ and $\left| {\Phi _{AP}^{\left( 2
\right)} } \right\rangle $ are two normalized states of the composite system
AP. $\left| {\bar {\Phi }_{AP}^{\left( i \right)} } \right\rangle $ is the
complement of $\left| {\Phi _{AP}^{\left( i \right)} } \right\rangle $.
Without loss of generality, the coefficients in Eq. (\ref{eq12}) and Eq. (\ref{eq13}) are
assumed to be positive real numbers. Obviously, Eq. (\ref{eq5}) and Eq. (\ref{eq6}) are
special case of Eq. (\ref{eq12}) and Eq. (\ref{eq13}) with $\left| {P^{\left( i \right)}}
\right\rangle = \left| {P_0 } \right\rangle $ and $\left| {\Phi
_{AP}^{\left( i \right)} } \right\rangle $ having a special decomposition.
We denote the subspace spanned by the state $\left| {P^{\left( i \right)}}
\right\rangle $ by the symbol $S_0 $. During the transformation, after the
unitary evolution a measurement of the probe with a postselection of the
measurement results projects its state into the subspace $S_0 $. After this
projection, the state of the system A should be $\frac{1}{\sqrt 2 }\left(
{\left| {\Psi _i } \right\rangle + \left| {\bar {\Psi }_i } \right\rangle }
\right)$ in Eq. (\ref{eq12}) and be $\frac{1}{\sqrt 2 }\left( {\left| {\Psi _i }
\right\rangle - \left| {\bar {\Psi }_i } \right\rangle } \right)$ in Eq.
(\ref{eq13}), so all the states $\left| {\Phi _{AP}^{\left( i \right)} }
\right\rangle $ ought to lie in a space orthogonal to $S_0 $. This requires

\begin{equation}
\label{eq14}
\left| {P^{\left( i \right)}} \right\rangle \left\langle {P^{\left( i
\right)}} \right|\left. {\Phi _{AP}^{\left( i \right)} } \right\rangle =
0,\left( {i = 1,2} \right).
\end{equation}

With the condition (\ref{eq14}), inter-inner-products of Eq. (\ref{eq12}) and Eq. (\ref{eq13}) yield
the following two matrix equations:

\begin{equation}
\label{eq15}
X^{\left( 1 \right)} = \sqrt \Gamma X_P^{\left( 2 \right)} \sqrt {\Gamma ^ +
} + \sqrt {I_n - \Gamma } Y_1 \sqrt {I_n - \Gamma ^ + } ,
\end{equation}

\begin{equation}
\label{eq16}
X^{\left( 3 \right)} = \sqrt \Lambda X_P^{\left( 4 \right)} \sqrt {\Lambda ^
+ } + \sqrt {I_n - \Lambda } Y_2 \sqrt {I_n - \Lambda ^ + } ,
\end{equation}

\noindent
where the $2\times 2$ matrix $Y_1 = \left[ {\left\langle {\Phi _{AP}^{\left(
i \right)} } \right|\left. {\Phi _{AP}^{\left( j \right)} } \right\rangle }
\right]$, $Y_2 = \left[ {\left\langle {\bar {\Phi }_{AP}^{\left( i \right)}
} \right|\left. {\bar {\Phi }_{AP}^{\left( j \right)} } \right\rangle }
\right]$, $X_P^{\left( 2 \right)} = \left[ {\frac{1}{2}\left( {\left\langle
{\Psi _i } \mathrel{\left| {\vphantom {{\Psi _i } {\Psi _j }}} \right.
\kern-\nulldelimiterspace} {\Psi _j } \right\rangle + \left\langle {\Psi _i
} \mathrel{\left| {\vphantom {{\Psi _i } {\bar {\Psi }_j }}} \right.
\kern-\nulldelimiterspace} {\bar {\Psi }_j } \right\rangle + \left\langle
{\bar {\Psi }_i } \mathrel{\left| {\vphantom {{\bar {\Psi }_i } {\Psi _j }}}
\right. \kern-\nulldelimiterspace} {\Psi _j } \right\rangle + \left\langle
{\bar {\Psi }_i } \mathrel{\left| {\vphantom {{\bar {\Psi }_i } {\bar {\Psi
}_j }}} \right. \kern-\nulldelimiterspace} {\bar {\Psi }_j } \right\rangle }
\right)\left\langle {P^{\left( i \right)}} \mathrel{\left| {\vphantom
{{P^{\left( i \right)}} {P^{\left( j \right)}}}} \right.
\kern-\nulldelimiterspace} {P^{\left( j \right)}} \right\rangle } \right]$
and $X_P^{\left( 4 \right)} = \left[ {\frac{1}{2}\left( {\left\langle {\Psi
_i } \mathrel{\left| {\vphantom {{\Psi _i } {\Psi _j }}} \right.
\kern-\nulldelimiterspace} {\Psi _j } \right\rangle - \left\langle {\Psi _i
} \mathrel{\left| {\vphantom {{\Psi _i } {\bar {\Psi }_j }}} \right.
\kern-\nulldelimiterspace} {\bar {\Psi }_j } \right\rangle - \left\langle
{\bar {\Psi }_i } \mathrel{\left| {\vphantom {{\bar {\Psi }_i } {\Psi _j }}}
\right. \kern-\nulldelimiterspace} {\Psi _j } \right\rangle + \left\langle
{\bar {\Psi }_i } \mathrel{\left| {\vphantom {{\bar {\Psi }_i } {\bar {\Psi
}_j }}} \right. \kern-\nulldelimiterspace} {\bar {\Psi }_j } \right\rangle }
\right)\left\langle {P^{\left( i \right)}} \mathrel{\left| {\vphantom
{{P^{\left( i \right)}} {P^{\left( j \right)}}}} \right.
\kern-\nulldelimiterspace} {P^{\left( j \right)}} \right\rangle } \right]$.
Following the proof of Lemma 2 in Ref. \cite{Duan:1998}, $Y_1 $ and
$Y_2 $ are positive-semidefinite matrixes, thus $\sqrt {I_n - \Gamma } Y_1
\sqrt {I_n - \Gamma ^ + } $ and $\sqrt {I_n - \Lambda } Y_2 \sqrt {I_n -
\Lambda ^ + } $ are positive-semidefinite matrixes. So $X^{\left( 1 \right)}
- \sqrt \Gamma X_P^{\left( 2 \right)} \sqrt {\Gamma ^ + } $ and $X^{\left( 3
\right)} - \sqrt \Lambda X_P^{\left( 4 \right)} \sqrt {\Lambda ^ + } $
should also be positive-semidefinite. On the other hand, if $X^{\left( 1
\right)} - \sqrt \Gamma X_P^{\left( 2 \right)} \sqrt {\Gamma ^ + } $ and
$X^{\left( 3 \right)} - \sqrt \Lambda X_P^{\left( 4 \right)} \sqrt {\Lambda
^ + } $ are positive-semidefinite matrixes, following the proof of Theorem
1, Eq. (\ref{eq15}) and Eq. (\ref{eq16}) can be satisfied with a special choice of $\left|
{\Phi _{AP}^{\left( i \right)} } \right\rangle $, then lemma 1 shows that
the transformation defined by Eq. (\ref{eq3}) could be implemented by a unitary
evolution together with a measurement such that for any unknown state chosen
from a set $A = \left\{ {\left| {\Psi _i } \right\rangle ,\left| {\bar {\Psi
}_i } \right\rangle } \right\}\left( {i = 1,2} \right)$. We thus get the
following theorem:

Theorem 2. There exists transformation defined by Eq. (\ref{eq3}) for any unknown
state chosen from a set $A = \left\{ {\left| {\Psi _i } \right\rangle
,\left| {\bar {\Psi }_i } \right\rangle } \right\}\left( {i = 1,2} \right)$
with diagonal efficiency matrix $\Gamma $ and $\Lambda $ if and only if the
matrixes $X^{\left( 1 \right)} - \sqrt \Gamma X_P^{\left( 2 \right)} \sqrt
{\Gamma ^ + } $ and $X^{\left( 3 \right)} - \sqrt \Lambda X_P^{\left( 4
\right)} \sqrt {\Lambda ^ + } $ are positive-semidefinite.

The semipositivity of the matrixes $X^{\left( 1 \right)} - \sqrt \Gamma
X_P^{\left( 2 \right)} \sqrt {\Gamma ^ + } $ and $X^{\left( 3 \right)} -
\sqrt \Lambda X_P^{\left( 4 \right)} \sqrt {\Lambda ^ + } $ give a series of
inequalities about the efficiencies $\gamma _i $ and $\delta _i $. The best
possible transformation efficiencies $\gamma _i $ and $\delta _i $ are
obtained by solving these inequalities. For example, if there are four
states $\left| {\Psi _1 } \right\rangle ,\left| {\Psi _2 } \right\rangle
,\left| {\bar {\Psi }_1 } \right\rangle ,$ and $\left| {\bar {\Psi }_2 }
\right\rangle $, we can show that the transformation efficiencies $\gamma $
and $\delta $ satisfy

\begin{equation}
\label{eq17}
\frac{\gamma _1 + \gamma _2 }{2} \le \frac{1 - \left| {\left\langle {\Psi _1
} \mathrel{\left| {\vphantom {{\Psi _1 } {\Psi _2 }}} \right.
\kern-\nulldelimiterspace} {\Psi _2 } \right\rangle } \right|}{1 -
\frac{1}{2}\left| {\left\langle {\Psi _1 } \mathrel{\left| {\vphantom {{\Psi
_1 } {\Psi _2 }}} \right. \kern-\nulldelimiterspace} {\Psi _2 }
\right\rangle + \left\langle {\Psi _1 } \mathrel{\left| {\vphantom {{\Psi _1
} {\bar {\Psi }_2 }}} \right. \kern-\nulldelimiterspace} {\bar {\Psi }_2 }
\right\rangle + \left\langle {\bar {\Psi }_1 } \mathrel{\left| {\vphantom
{{\bar {\Psi }_1 } {\Psi _2 }}} \right. \kern-\nulldelimiterspace} {\Psi _2
} \right\rangle + \left\langle {\bar {\Psi }_1 } \mathrel{\left| {\vphantom
{{\bar {\Psi }_1 } {\bar {\Psi }_2 }}} \right. \kern-\nulldelimiterspace}
{\bar {\Psi }_2 } \right\rangle } \right|},
\end{equation}

\begin{equation}
\label{eq18}
\frac{\delta _1 + \delta _2 }{2} \le \frac{1 - \left| {\left\langle {\bar
{\Psi }_1 } \mathrel{\left| {\vphantom {{\bar {\Psi }_1 } {\bar {\Psi }_2
}}} \right. \kern-\nulldelimiterspace} {\bar {\Psi }_2 } \right\rangle }
\right|}{1 - \frac{1}{2}\left| {\left\langle {\Psi _1 } \mathrel{\left|
{\vphantom {{\Psi _1 } {\Psi _2 }}} \right. \kern-\nulldelimiterspace} {\Psi
_2 } \right\rangle - \left\langle {\Psi _1 } \mathrel{\left| {\vphantom
{{\Psi _1 } {\bar {\Psi }_2 }}} \right. \kern-\nulldelimiterspace} {\bar
{\Psi }_2 } \right\rangle - \left\langle {\bar {\Psi }_1 } \mathrel{\left|
{\vphantom {{\bar {\Psi }_1 } {\Psi _2 }}} \right.
\kern-\nulldelimiterspace} {\Psi _2 } \right\rangle + \left\langle {\bar
{\Psi }_1 } \mathrel{\left| {\vphantom {{\bar {\Psi }_1 } {\bar {\Psi }_2
}}} \right. \kern-\nulldelimiterspace} {\bar {\Psi }_2 } \right\rangle }
\right|},
\end{equation}

For the state chosen from polar great circle, we have

\[
\left\langle {\Psi _1 } \mathrel{\left| {\vphantom {{\Psi _1 } {\Psi _2 }}}
\right. \kern-\nulldelimiterspace} {\Psi _2 } \right\rangle =
\frac{1}{2}\left( {\left\langle {\Psi _1 } \mathrel{\left| {\vphantom {{\Psi
_1 } {\Psi _2 }}} \right. \kern-\nulldelimiterspace} {\Psi _2 }
\right\rangle + \left\langle {\Psi _1 } \mathrel{\left| {\vphantom {{\Psi _1
} {\bar {\Psi }_2 }}} \right. \kern-\nulldelimiterspace} {\bar {\Psi }_2 }
\right\rangle + \left\langle {\bar {\Psi }_1 } \mathrel{\left| {\vphantom
{{\bar {\Psi }_1 } {\Psi _2 }}} \right. \kern-\nulldelimiterspace} {\Psi _2
} \right\rangle + \left\langle {\bar {\Psi }_1 } \mathrel{\left| {\vphantom
{{\bar {\Psi }_1 } {\bar {\Psi }_2 }}} \right. \kern-\nulldelimiterspace}
{\bar {\Psi }_2 } \right\rangle } \right)
\]

\begin{equation}
\label{eq19}
 = \left\langle {\bar {\Psi }_1 } \mathrel{\left| {\vphantom {{\bar {\Psi
}_1 } {\bar {\Psi }_2 }}} \right. \kern-\nulldelimiterspace} {\bar {\Psi }_2
} \right\rangle = \frac{1}{2}\left( {\left\langle {\Psi _1 } \mathrel{\left|
{\vphantom {{\Psi _1 } {\Psi _2 }}} \right. \kern-\nulldelimiterspace} {\Psi
_2 } \right\rangle - \left\langle {\Psi _1 } \mathrel{\left| {\vphantom
{{\Psi _1 } {\bar {\Psi }_2 }}} \right. \kern-\nulldelimiterspace} {\bar
{\Psi }_2 } \right\rangle - \left\langle {\bar {\Psi }_1 } \mathrel{\left|
{\vphantom {{\bar {\Psi }_1 } {\Psi _2 }}} \right.
\kern-\nulldelimiterspace} {\Psi _2 } \right\rangle + \left\langle {\bar
{\Psi }_1 } \mathrel{\left| {\vphantom {{\bar {\Psi }_1 } {\bar {\Psi }_2
}}} \right. \kern-\nulldelimiterspace} {\bar {\Psi }_2 } \right\rangle }
\right),
\end{equation}

\noindent
under the condition Eq. (\ref{eq19}), Eq. (\ref{eq17}) and Eq. (\ref{eq18}) reduce to $\frac{\gamma
_1 + \gamma _2 }{2} \le 1$ and $\frac{\delta _1 + \delta _2 }{2} \le 1$,
then the transformation efficiencies $\gamma $ and $\delta $ could both
reach unity which accords with the results of Ref. \cite{Pati:2002}.

In the following we will show how to implement the transformation defined by
Eq. (\ref{eq4}). The result is the following theorem:

Theorem 3. There exists a unitary operator $U$ such that for any unknown
state chosen from a set $A = \left\{ {\left| {\Psi _i } \right\rangle
,\left| {\bar {\Psi }_i } \right\rangle } \right\}\left( {i = 1,2} \right)$

\begin{equation}
\label{eq20}
U\left( {\left| {\Psi _i } \right\rangle \left| {P_0 } \right\rangle }
\right) = \sqrt {\gamma _i } \frac{1}{\sqrt 2 }\left( {a\left| {\Psi _i }
\right\rangle + b\left| {\bar {\Psi }_i } \right\rangle } \right)\left| {P_0
} \right\rangle + \sum\limits_{j = 1}^2 {a_{ij} \left| {\Phi _A^{\left( j
\right)} } \right\rangle } \left| {P_j } \right\rangle ,
\end{equation}

\begin{equation}
\label{eq21}
U\left( {\left| {\bar {\Psi }_i } \right\rangle \left| {P_0 } \right\rangle
} \right) = \sqrt {\delta _i } \frac{1}{\sqrt 2 }\left( {b^\ast \left| {\Psi
_i } \right\rangle - a^\ast \left| {\bar {\Psi }_i } \right\rangle }
\right)\left| {P_0 } \right\rangle + \sum\limits_{j = 1}^2 {b_{ij} \left|
{\bar {\Phi }_A^{\left( j \right)} } \right\rangle } \left| {P_j }
\right\rangle ,
\end{equation}

\noindent
for some real numbers $\gamma _i \ge 0$, $\delta _i \ge 0$, $a_{ij} \ge 0$,
and $b_{ij} \ge 0$ with $i,j = 1,2$, if and only if $\left| {\Psi _1 }
\right\rangle ,\left| {\Psi _2 } \right\rangle ,\left| {\bar {\Psi }_1 }
\right\rangle ,\left| {\bar {\Psi }_2 } \right\rangle $ are linearly
independent. The inter-inner-products of Eq. (\ref{eq20}) and Eq. (\ref{eq21}) yield the
following two matrix equations

\begin{equation}
\label{eq22}
X^{(\ref{eq1})} = \sqrt \Gamma X^{\left( 5 \right)}\sqrt {\Gamma ^ + } + AA^ + ,
\end{equation}

\begin{equation}
\label{eq23}
X^{(\ref{eq3})} = \sqrt \Lambda X^{\left( 6 \right)}\sqrt {\Lambda ^ + } + BB^ + ,
\end{equation}

\noindent
where the $2\times 2$ matrixes $X^{\left( 5 \right)} = \left[ {\left| a
\right|^2\left\langle {\Psi _i } \mathrel{\left| {\vphantom {{\Psi _i }
{\Psi _j }}} \right. \kern-\nulldelimiterspace} {\Psi _j } \right\rangle +
a^\ast b\left\langle {\Psi _i } \mathrel{\left| {\vphantom {{\Psi _i } {\bar
{\Psi }_j }}} \right. \kern-\nulldelimiterspace} {\bar {\Psi }_j }
\right\rangle + ab^\ast \left\langle {\bar {\Psi }_i } \mathrel{\left|
{\vphantom {{\bar {\Psi }_i } {\Psi _j }}} \right.
\kern-\nulldelimiterspace} {\Psi _j } \right\rangle + \left| b
\right|^2\left\langle {\bar {\Psi }_i } \mathrel{\left| {\vphantom {{\bar
{\Psi }_i } {\bar {\Psi }_j }}} \right. \kern-\nulldelimiterspace} {\bar
{\Psi }_j } \right\rangle } \right]$, and $X^{\left( 6 \right)} = \left[
{\left| a \right|^2\left\langle {\Psi _i } \mathrel{\left| {\vphantom {{\Psi
_i } {\Psi _j }}} \right. \kern-\nulldelimiterspace} {\Psi _j }
\right\rangle - a^\ast b\left\langle {\Psi _i } \mathrel{\left| {\vphantom
{{\Psi _i } {\bar {\Psi }_j }}} \right. \kern-\nulldelimiterspace} {\bar
{\Psi }_j } \right\rangle - ab^\ast \left\langle {\bar {\Psi }_i }
\mathrel{\left| {\vphantom {{\bar {\Psi }_i } {\Psi _j }}} \right.
\kern-\nulldelimiterspace} {\Psi _j } \right\rangle + \left| b
\right|^2\left\langle {\bar {\Psi }_i } \mathrel{\left| {\vphantom {{\bar
{\Psi }_i } {\bar {\Psi }_j }}} \right. \kern-\nulldelimiterspace} {\bar
{\Psi }_j } \right\rangle } \right]$£¬ the following procedure of verifying
theorem 3 is similar to that of Theorem 1, and we omit the details. Within
the same approach of verifying Theorem 2, we have the following corollary:

Corollary. There exists transformation defined by Eq. (\ref{eq4}) for any unknown
state chosen from a set $A = \left\{ {\left| {\Psi _i } \right\rangle
,\left| {\bar {\Psi }_i } \right\rangle } \right\}\left( {i = 1,2} \right)$
with diagonal efficiency matrixes $\Gamma $ and $\Lambda $ if and only if
the matrixes $X^{(\ref{eq1})} - \sqrt \Gamma X^{\left( 5 \right)}\sqrt {\Gamma ^ + }
$ and $X^{(\ref{eq3})} - \sqrt \Lambda X^{\left( 6 \right)}\sqrt {\Lambda ^ + } $
are positive-semidefinite.

It should be pointed out that our probabilistic quantum
information processing machine could also be comprehended in the
concept of quantum program\cite{Nielsen:1997}, which was
originally discussed by Nielsen and Chuang \cite{Nielsen:1997}.
The theory of quantum program has been developed in Ref.
\cite{Vidal:2002,Hillery:2001,Hillery:2002,Hillery:2003,Dusek:2002,Hillery:2004}.
In Ref. \cite{Hillery:2004} M. Hillery et al show that an
arbitrary SU(\ref{eq2}) transformation of qubits can be encoded in
program state of a universal programmable probabilistic quantum
processor. The probability of success of this processor can be
enhanced by systematic correction of errors via conditional loops.
Our probabilistic quantum information processing machine could
also be regarded as a special probabilistic programmable quantum
processor. There are two inputs of this machine: a data state,
which is chosen from a set $A = \left\{ {\left| {\Psi _i }
\right\rangle ,\left| {\bar {\Psi }_i } \right\rangle }
\right\}\left( {i = 1,2} \right)$, and a program state. In our
machine, the program state is represented by the probe $P$. The
instructions of the quantum program are to perform the Hadamard
and Unitary gates defined by Eq. (\ref{eq3}) and Eq. (\ref{eq4})
for the input qubits. The transformation efficiencies of our
probabilistic quantum information processing machines dependent on
the value of $a$ and $b$, whereas the probabilistic quantum
processor constructed in Ref. \cite{Hillery:2004} has the same
transformation efficiencies for arbitrary SU(\ref{eq2}) rotation.
So we conclude that our machine is not universal for arbitrary
unitary operation, but it might reach higher transformation
efficiencies for some special unitary operation compared to the
probabilistic quantum processor constructed in Ref.
\cite{Hillery:2004}.

To conclude, we have constructed some probabilistic quantum information
processing machines to implement the Hadamard and Unitary gates defined by
Eq. (\ref{eq3}) and Eq. (\ref{eq4}). As we know, these gates are very useful in various
quantum algorithms and information processing protocols, while linearity of
quantum mechanics does not allow linear superposition of an unknown qubit
with its complement. Our results show that these quantum operations can be
implemented perfectly with non-zero probabilities. The best transformation
efficiencies are derived. We expect our results to play a fundamental role
in future understanding of quantum information theory. There still leaves an
interesting question that is how to construct an approximately universal
machine to realize the transformation defined by Eq. (\ref{eq3}) and Eq. (\ref{eq4}) in a
deterministic way.

{\bf Acknowledgement}

We thank Yong-Sheng Zhang for helpful discussions and suggestions. This work
is supported by Anhui Provincial Natural Science Foundation under Grant No:
03042401, the Key Program of the Education Department of Anhui Province
under Grant No: 2004kj005zd and the Talent Foundation of Anhui University.

\end{document}